\documentclass[conference]{IEEEtran}
\IEEEoverridecommandlockouts
\usepackage{cite}
\usepackage{amsmath,amssymb,amsfonts}
\usepackage{algorithmic}
\usepackage{graphicx}
\usepackage{textcomp}
\usepackage{xcolor}
\def\BibTeX{{\rm B\kern-.05em{\sc i\kern-.025em b}\kern-.08em
    T\kern-.1667em\lower.7ex\hbox{E}\kern-.125emX}}
\begin{document}

\title{How advertising strategies affects the diffusion of information in markets*\\
{\footnotesize \textsuperscript{*}Note: This is only predraft of my thesis and not submitted}
\thanks{Codes is available on github.}
}

\author{\IEEEauthorblockN{\textsuperscript{} Eren Arkangil}
\IEEEauthorblockA{\textit{Computer Engineering} \\
\textit{Bogazici University}}

}

\maketitle

\begin{abstract}
Consumer behavior under social influence is a well-known phenomenon and computer scientists and economists are prevalently trying to analyze the dynamics behind decision making during the consumption process through agent-based modeling (ABM). Some articles tried to explain market inequality~\cite{herdagdelen2008cultural} because of the social influencing, but the impact of advertising is underestimated and not included as a parameter in the ABM simulations. In the first part of the work we give a background about related works, afterwards, we explain our model with newly introduced advertisement and penalty parameters. To best our knowledge our work will be the first paper that will consider the effects of social influencing, advertisement, and the novelty of the product at the same time. The type of interactions is defined as advertisement and social influencing. We are interested in showing the effects of advertisement and social interactions in different time intervals. The influencing takes time by its nature, however, advertisement is a stronger approach to introduce new products to consumers. These effects are not linearly positive for the fashion products, since the fashion changes over time and consumed items get old-fashioned and ordinary for users. Our Sigmoid penalty function adds non-linearity to the model to show the time and popularity effects against the advertisement and social interactions.

\end{abstract}
\tableofcontents
%



\section{INTRODUCTION}
\label{chapter:INTRODUCTION}
A vast amount of fashion products are provided to millions on e-commerce sites. Purchasing the most suitable one between almost unlimited options is a very challenging and time-consuming process. The capacity of human memory is constrained to make such decision ~\cite{simon1974big}. Many computational models and computer simulations do not take into consideration the irrational behavior of individuals due to limited time and knowledge ~\cite{gigerenzer1996reasoning}. Generally, firms try to take advantage of this situation and to sell as many products as possible by following different advertisement strategies. In this study, we will investigate the impact of advertisements on people’s consumption decisions. We present a fashion market model and the results we obtain from simulations are going to be existence proof that following the most aggressive advertisement is necessarily not the most rational strategy.

\section{RELATED WORKS}
\label{chapter:METHODOLOGY}
The dynamics of decision making is still an open question to debate and some models focus on the effect of social influence on the decisions, while some others assert that it is effected by advertisement ~\cite{ herdagdelen2008cultural, cetin2014attention}. The simulations made in these articles point outs that people are not perfectly rational in their consumption preferences due to lack of information. Information plays a crucial role in consumption decisions~\cite{stuart2016exploring}. Social influencing can be explained as a information exchange between individual agents. The agent-based simulations on graphs are commonly used in some articles~\cite{ herdagdelen2008cultural, cetin2014attention} to analyze social influencing between the members of a society. Graph is the data type used to conceptualize the structure of the society. A graph $G = (N,E) $ consist of edges and nodes. $N$ is a set of nodes. A graph may have different type of topologies. For example, ring lattice is used before to explain the structure of the society~\cite{herdagdelen2008cultural}. Small-world graph is an another graph type used to explore miscellaneous fields from cultural networks~\cite{uzzi2005collaboration} or social influence networks~\cite{kitsak2010identification} to the airport networks~\cite{guida2007topology} and semantic networks~\cite{beckage2011small}

\subsection{A Cultural Market}

The concept of the market in economics signifies an environment, in which buyers and sellers can exchange different types of goods and information. Therefore, a market consists of consumers from the point of sellers and goods from the point of buyers. In our article, we name consumers as agents. The goods on sale  for the agents in the market are named as items. A Cultural market is a special type of market that agents are interested in consuming cultural items such as books and movies. One of the assumption made in the cultural market model~\cite{herdagdelen2008cultural} that the market is a perfect competition.Perfect competition urges that all of the agents in the market have perfect information about the items. It means that all agents have perfect knowledge of the quality of the items in the market. The perfect information is a key assumption of neo-classical economics~\cite{leahey2003herbert}
Information asymmetry term in behavioral economics is still questioning whether the occurrence of perfect information in the market possible. If one of the agents has better information than the other, information asymmetry occurs. This consequence of information asymmetry might cause market inefficiency in the consumption distribution of the items in the market according to their quality. Market inequality due to the communication process of the agents is still an active research topic. 
Fashion market is a market that consists of fashion items such as clothes or t-shirts. Fashion markets are ‘’fast” ~\cite{bhardwaj2010fast,sull2008fast}. The expected agent behavior is to keep consuming new fashion items. According to the reports it takes $4-8$ weeks that people get bored of a fashion item~\cite{hines2001globalization}. We are interested in discovering the consumer behavior in according to marketing strategy of the producer.

\section{METHODOLOGY}
\label{chapter:OBJECTIVES}
The cultural market model explained above have 4 parameters and 2 quality measurement metrics that needs to be defined before we proceed to extend this model to the fashion market.
\subsection{Cultural Market Model}
The liking value $L_{i \alpha}$, is the preference of the agent $i$ on  the item $\alpha$. $L_{i \alpha} $can take values between $0$ and $1$. Each agent sort its preferences on each item according to their tastes. This taste determines the value of the $L_{i \alpha}$, most favorable item of an agent takes $1$ and the least takes $0$. Agents have different taste from each other. To our best knowledge, there is no publicly available dataset, therefore we are setting this values from an uniform distrubition. This values are set once and same during the simulation.
The social pressure parameter $S_{i \alpha}$ represents the impact of social influencing on agent $i$ to consume item $\alpha$. It is calculated by the ratio of the number of neighbors of agent $i$, who consumed item a before. The lambda value $\gamma$  determines how much an agent is effected by the information he gets from the neighbor agent.
The opinion $O_{i \alpha}$ is the weighted average of $S_{i \alpha}$ and $L_{i \alpha}$. It is time independent value unlike $L_{i \alpha}$, and changes each turn. The agents rank the items corresponding to $O_{i \alpha}$ values each turn and consume the item with the highest $O_{i \alpha}$ value. One item only can be consumed once per agent.
The items have different qualities. Quality $Q_{ \alpha}$ is calculated as the expected average value of the liking values of an item over all agents. \[Q_{ \alpha} = \sum\limits_{i=1}^N  L_{i \alpha} /N\]
The market share $C_{ \alpha}$ how many times the item a has been consumed by agents.It is calculated as $C_{ \alpha} =\sum\limits_{i=1}^N C_{i \alpha} /N $ where $C_{i\alpha}$ is $1$ if agent i has consumed the item $\alpha$ before the current time step and to $0$ otherwise~\cite{herdagdelen2008cultural}. The market inequality represents the difference between the market shares of the cultural items. It is shown as a result of the irrational behavior of agents in the market due to limited information capability.
All in all, the model explained here is only covers cultural items and the results might be biased.Therefore, the model and simulations should be further investigated for a different type of markets.

\subsection{Fashion Market Model}
We propose a new model as an extension of the cultural market model~\cite{herdagdelen2008cultural} to explore the effect of advertisement on the market share of fashion items.This model takes into consideration of the depreciation effect of old fashioned goods and introduces a penalty parameter. In response companies introduce new items to the market for providing constant agent satisfaction, thereby try to protect their market shares and compete with each other to make the most sale. In the cultural market the market size was limited with $M$, however in our model firms introduce new items to the market after $6$ rounds.
The information asymmetry problem is discussed in the cultural market in the context of social influencing and it was the only information which flows through agents. The liking values of the new items will be $0$ since they are unknown and agents are not able to form their opinion about them.

\subsection{New Items and Advertisement}
Social influencing takes time to show its effect on consumption decisions because the consumers need to buy an item to recommend it. Thus, influencing as the only information source is not sufficient to analyze a market such fashion. People generally take into consideration the advertisement in their fashion decisions~\cite{leahey2003herbert,stuart2016exploring}.The advertisement will be the only information source for the new items until enough agents consume the new items and activate social pressure parameter $S_{i \alpha}$. We set a new parameter advertisement $A_{ \alpha}$ to show the effect of advertisement on new items. Advertisement is a necessary tool for firms to introduce their product to the market and the amount of it should be decided by its firms according to their budget. 
Some companies are able to follow aggressive advertisement, while others are content with modest ones. $A_{ \alpha}$ value corresponds to the \emph{advertisement strategy} of item  $\alpha$ by the firm. It can take values between $0$ and $1$. If its value is $1$, it means the firm decided to follow the most aggressive strategy, and $0$ stands for there is no advertisement.
However each agent reacts to the advertisement differently. Hence, we introduce the \emph{tolerance parameter} $T_{i}$. This parameter affects only agents and the value of $T$ is the same for every item. $T_{i}$ determines the vulnerability of agents to the advertisement. It must take continuous numeric values from strictly greater than zero and until $1$. The agents with higher tolerance values are affected by advertisement considerably more.
As a conclusion, We introduce the \emph{marketing parameter} $M_{i \alpha}$, which determines the preference of an agent on new item $\alpha$. $M_{i\alpha}$ values are not changing in time and only set once. It solves the cold-start problem of new introduced items in particular. The  $M_{i\alpha}$ represents the effect of the advertisement of the item $\alpha$ on agent $i$. It is calculated by the multiplication the advertisement and tolerance values. Higher values point out that the agent is more likely to consume this item rather than the other item with a lower value. 

\[
 	 M_{i \alpha} = A_{ \alpha}  T_{i}
\]

\subsection{Sigmoid Penalty Function}
The penalty function is a cumulative distribution function that will penalize an item if it is consumed too much. Excess of consumption is a sign of an overly commercialized item. We expect it is going to lose its popularity soon and people will lose interest to buy this item. 
\emph{Penalty} $P_{ \alpha}$ is applied to each agent in the same amount. We assume that it is common sense and affects every agent in the same amount. One of the variables of the penalty function is $f$. It is calculated with the inverse logit, alias sigmoid function. The sigmoid function is commonly used in marketing research for logit analysis of an item. It attempts to predict the purchase interest of an agent for the new item in the market~\cite{flath1979comparison,johansson1979advertising}.We will scale the x-axis of the sigmoid function between $0-1$, which will represent the ratio of the population who consumed item $\alpha$. Therefore the penalty coefficient $p$ is calcualted by \(
	f_{x} 
	= \frac{1}{1 + e^{-x}}
\). $X$ represents $C_{ \alpha}$ for this function.

\begin{figure}[htbp]
\begin{center}
	\includegraphics[width=\columnwidth]
		{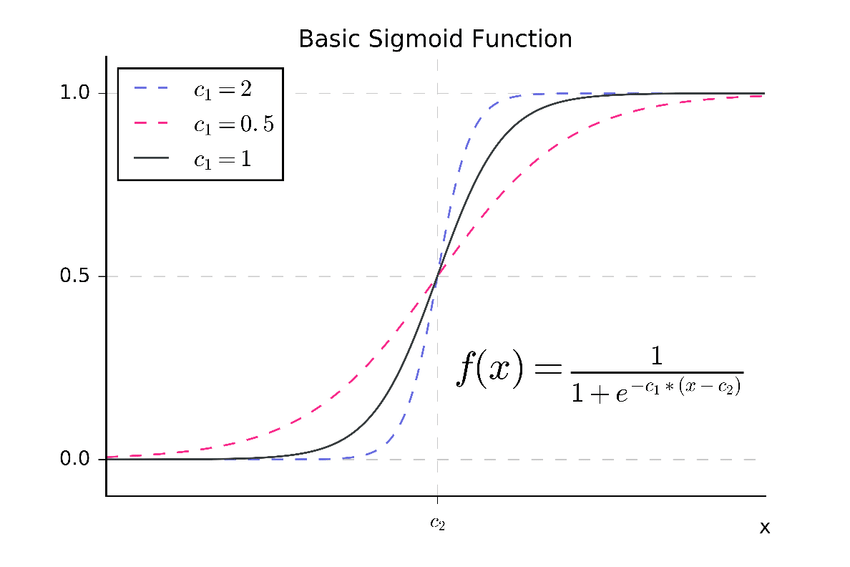}
	\caption{Sigmoid Function}
	\label{fig:sgmd}
\end{center}
\end{figure}

In the cultural market, the agents had opinion and preferences parameters, and the items were homogeneous where agents consider types of items to be identical and have no superiority over each other. In our model, the items are heterogeneous in regard to their advertisement values. Besides, items are penalized differently according to their advertisement strategy. Extremely aggresive advertisement should be penalized . Thus, there should be correlation between advertisement and consumption. We set our penalization function as 
\[
  P_{ \alpha} = f_{ \alpha}  A_{ \alpha}
\]
The sharpness of the sigmoid function can depend on other parameters such as socio-ecological factors, but for simplicity, we only use market share in this equation~\cite{flath1979comparison}.

\subsection{Experimental Design}
The aggressive advertisement strategy clearly ends up with fast selling, 
although this leadership won't last long on account of penalty function. We expect a moderate advertisement strategy would be optimal 
since it won't get penalized too much and provide enough information to agents to buy the new product.  
We will calculate the market share of the new and advertised items at each turn of simulation and monitor the pattern of the advertised item. 
Since the pleasure of consuming old items reduces thanks to penalty parameter as the simulation proceeds, 
the firms should give a decision what is the optimal marketing strategy to keep their product favorable as long as possible. 
We model this similar to the opinion parameter in the cultural market model. The utility $U_{i \alpha}$ shows the pleasure of consuming and fashion item and calculated each turn. The firms try to provide maximum $U_{i \alpha}$ to the agents with minimum advertisement value, 
since aggresive advertisement is costly for the firms and if the item gets too commercialized, the sales will drop.

\begin{align}
  U_{i \alpha} = 
                      { \gamma} S_{i \alpha} + (1-{ \gamma}) C_{i \alpha} + M_{i \alpha} - P_{ \alpha}
	\label{eq:eq1}
\end{align}

In the simulation, agents will compare new items with old items by ranking function $U_{i \alpha}$. Its a time-dependent variable and will be used to rank the items.
As the old items lose their popularity, the new ones will be popular in the market. A strong, but not the maximum advertisement strategy would be durable and robust.

\section{Results}
\label{chapter:Results}
All results reported in this section are based on values obtained by averaging over
100 independent runs of the simulation with the same parameters. It is natural to ask if the quality of an item determines its market share at the
end or not. A reasonable expectation about a fashion market is that items with
high advertisement values should get higher market shares on the average. We keep the number of agents and the number of items fixed to 100 in this set of
experiments (i.e. N = 30 and M = 50). We let the model run for
5, 20 and 30 steps. 
\begin{figure}[htbp]
\begin{center}
	\includegraphics[width=\columnwidth]
		{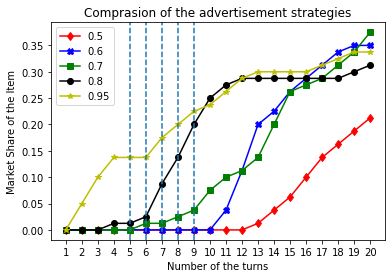}
	\caption{s2}
	\label{fig:sgmd}
\end{center}
\end{figure}
In Fig. 3, we see the scatter plot of advertisement
versus market shares of the items for different advertisement values. As we already noted before, the penalty parameter (depends on socioecological factors) is an important parameter and should be set carefully with
keeping in mind the actual market to be modeled. It is obvious that values of P very
close to extreme values are highly unrealistic. We set p different numbers during the next set of experiments for simplicity of analyses
\begin{figure}[htbp]
\begin{center}
	\includegraphics[width=\columnwidth]
		{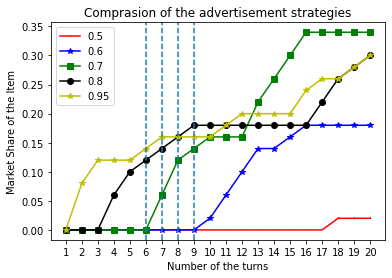}
	\caption{s3}
	\label{fig:sgmd}
\end{center}
\end{figure}
An interesting finding that, the aggresive advertisement strategy ends up with the fast increase rate and the advertised items dominates the market for a while, however, once an item is too commercialized \cite{stuart2016exploring} it loses its attraction to the consumers . On the other hand, if the advertisement is too low, it takes too much time to rise in the market and the share gained is comperatively low unlike moderate advertisement values.

\begin{figure}[htbp]
\begin{center}
	\includegraphics[width=\columnwidth]
		{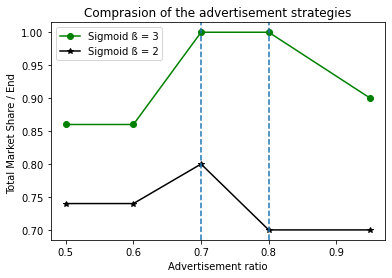}
	\caption{s4}
	\label{fig:sgmd}
\end{center}
\end{figure}

We run the experiments with the different $\beta$ valuess, which represent behaviour of the society against advertised items. As the beta values increases, the people of this society is more harsh to the advertised items.

\section{ Conclusions and Future Work}

Empirical findings suggested that the introduction of advertisement has a profound effect on the market share of an item. For
high values of advertisement   i.e. $(y = 0.95)$ , we observed that as we increase the
advertisement parameter $(a = 0.7)$, the linearity was disrupted and we observed
nonlinear relation in favor of moderately advertised items.

We carried out extended simulations to see if our results depend on specific values
of the parameters or robust to different values of the parameters. We concluded
that the qualitative nature of the simulations are robust with respect to different
number of agents (i.e. $ N \rho {100, 500, 1000, 5000})$ and varying degrees of network
connectivity 

We studied the ring topology as a base model and introduced the heterogeneity
by using random topology. How the model will behave if another type of network
such as scale free or small world is introduced is definitely an interesting and nontrivial question and needs to be addressed in the future studies. Also the effect of
introducing directed links and asymmetrical neighboring relations can be investigated in the future.


\bibliographystyle{unsrt}
\bibliography{references}

\begin{thebibliography}{10}

\bibitem{herdagdelen2008cultural}
Ama{\c{c}} Herdaǧdelen and Haluk Bingol.
\newblock A cultural market model.
\newblock {\em International Journal of Modern Physics C}, 19(02):271--282,
  2008.

\bibitem{simon1974big}
Herbert~A Simon.
\newblock How big is a chunk?: By combining data from several experiments, a
  basic human memory unit can be identified and measured.
\newblock {\em Science}, 183(4124):482--488, 1974.

\bibitem{gigerenzer1996reasoning}
Gerd Gigerenzer and Daniel~G Goldstein.
\newblock Reasoning the fast and frugal way: models of bounded rationality.
\newblock {\em Psychological review}, 103(4):650, 1996.

\bibitem{cetin2014attention}
Uzay Cetin and Haluk~O Bingol.
\newblock Attention competition with advertisement.
\newblock {\em Physical Review E}, 90(3):032801, 2014.

\bibitem{stuart2016exploring}
Jillian~O'Rourke Stuart.
\newblock Exploring the impact of power on information consumption decisions.
\newblock 2016.

\bibitem{uzzi2005collaboration}
Brian Uzzi and Jarrett Spiro.
\newblock Collaboration and creativity: The small world problem.
\newblock {\em American journal of sociology}, 111(2):447--504, 2005.

\bibitem{kitsak2010identification}
Maksim Kitsak, Lazaros~K Gallos, Shlomo Havlin, Fredrik Liljeros, Lev Muchnik,
  H~Eugene Stanley, and Hern{\'a}n~A Makse.
\newblock Identification of influential spreaders in complex networks.
\newblock {\em Nature physics}, 6(11):888--893, 2010.

\bibitem{guida2007topology}
Michele Guida and Funaro Maria.
\newblock Topology of the italian airport network: A scale-free small-world
  network with a fractal structure?
\newblock {\em Chaos, Solitons \& Fractals}, 31(3):527--536, 2007.

\bibitem{beckage2011small}
Nicole Beckage, Linda Smith, and Thomas Hills.
\newblock Small worlds and semantic network growth in typical and late talkers.
\newblock {\em PloS one}, 6(5):e19348, 2011.

\bibitem{leahey2003herbert}
Thomas~H Leahey.
\newblock Herbert a. simon: Nobel prize in economic sciences, 1978.
\newblock {\em American Psychologist}, 58(9):753, 2003.

\bibitem{bhardwaj2010fast}
Vertica Bhardwaj and Ann Fairhurst.
\newblock Fast fashion: response to changes in the fashion industry.
\newblock {\em The international review of retail, distribution and consumer
  research}, 20(1):165--173, 2010.

\bibitem{sull2008fast}
Donald Sull and Stefano Turconi.
\newblock Fast fashion lessons.
\newblock {\em Business Strategy Review}, 19(2):4--11, 2008.

\bibitem{hines2001globalization}
Tony Hines.
\newblock Globalization: An introduction to fashion markets and fashion
  marketing.
\newblock {\em Fashion marketing: Contemporary issues}, pages 121--32, 2001.

\bibitem{flath1979comparison}
David Flath and EW~Leonard.
\newblock A comparison of two logit models in the analysis of qualitative
  marketing data.
\newblock {\em Journal of Marketing Research}, 16(4):533--538, 1979.

\bibitem{johansson1979advertising}
Johny~K Johansson.
\newblock Advertising and the s-curve: A new approach.
\newblock {\em Journal of Marketing Research}, 16(3):346--354, 1979.

\end{thebibliography}

\end{document}